# The Doppler effect from a uniformly moving mirror

Aleksandar Gjurchinovski

*Department of Physics, Faculty of Natural Sciences and Mathematics,*
*Sts. Cyril and Methodius University,*
*P. O. Box 162, 1000 Skopje, Macedonia*

Electronic mail: agjurcin@iunona.pmf.ukim.edu.mk, agjurcin@yahoo.com

**ABSTRACT**

The formula for the frequency shift of a plane-polarized light wave reflected from a uniformly moving mirror is derived directly from the constant light speed postulate and the basic principles of wave optics. Unlike the original derivation by Einstein, our derivation does not involve Lorentz transformations. As such, it lies within the scope of a first course in special relativity.

A plane-polarized light wave reflected from a uniformly moving mirror will suffer a frequency change [1]. This special case of the Doppler effect was predicted and described by Einstein in his famous 1905 paper on special relativity [2]. To show that the value of this frequency shift depends on the angle of incidence of the wave and the velocity of the moving mirror, Einstein employed Lorentz transformations on the equations describing the reflection in the reference frame where the mirror was at rest. It is interesting to note that this Doppler phenomenon is a particular case of a special-relativistic scattering by a moving object, a subject vastly researched in the last few decades [3,4,5].

The purpose of the present paper is to derive the formula for the Doppler effect from a uniformly moving mirror directly from the constant light speed postulate and the basic principles of wave optics. Unlike the original derivation by Einstein, our derivation does not involve Lorentz transformations. As such, it lies within the scope of a first course in special relativity.

Figure 1 shows a plane-polarized light wave incident on a vertical plane mirror. The mirror is moving at a constant velocity $v$ to the right. We denote by $\alpha$ the angle of incidence, and by $\beta$ the angle of reflection of the light.[1] We let $A$ and $B$ to be the points of two successive wavefronts of the incident light a distance $\overline{AB} = \lambda_0$ apart, where $\lambda_0$ equals the wavelength of the incident light. The wavefront at point $B$ will be reflected by the mirror at time $t_0$. On the other hand, the wavefront at $A$ will have to travel the path $\overline{AA'} = c(t - t_0)$ before being reflected by the mirror. The reflection will occur off the point $A'$ at time $t$, when the mirror is located a distance $v(t - t_0)$ to the right from its position at time $t_0$. For this same amount of time $(t - t_0)$ required for the wavefront at $A$ to cover the path $\overline{AA'}$, the wavefront reflected at $B$ will traverse the path $\overline{BB'} = c(t - t_0)$. We have taken into account the constant light speed postulate, according to which the speed of the reflected wavefront will equal $c$, and, thus, will not depend on the motion of the reflecting surface [6,7].

Observe from Fig. 1 that the paths of the reflected wavefronts will not coincide due to the shift of the wavefronts caused by the motion of the mirror [6]. Also observe that the distance $\overline{A''B'}$ is the shortest (orthogonal) distance between the planes of the reflected wavefronts located at $A'$ and $B'$, and it equals the wavelength $\lambda$ of the reflected light. From Fig. 1 we have

$$\overline{AA'} = \overline{AB} + \overline{BA'}, \tag{1}$$

and

$$\overline{A''B'} = \overline{A''B} + \overline{BB'}. \tag{2}$$

---

[1] At this point, we made an implicit assumption that the usual law of reflection would not hold in the case of a uniformly moving mirror, that is, the angles of incidence $\alpha$ and reflection $\beta$ of the light would not equal to each other. This well-known fact has been emphasized for the first time also by Einstein, and also in his celebrated 1905 paper [2]. Einstein derived the law of reflection of light from a uniformly moving mirror in the same manner in which he obtained the Doppler formula from a uniformly moving mirror, that is, by employing the standard Lorentz-transformation procedure to switch from the reference frame where the mirror is stationary to the reference frame where the mirror is in uniform rectilinear motion. The formula for the law of reflection from a uniformly moving mirror also can be obtained directly from the constant light speed postulate, by using a simple Huygens' construction [6] or Fermat's principle of least time [7]. Another interesting way to derive the reflection formula is given in the paper by Censor [3].

From the triangles $A'BC$ and $A'A''B$, we have

$$\overline{BA'} = \frac{v(t-t_0)}{\cos\alpha}, \tag{3}$$

and

$$\overline{A''B} = \overline{BA'}\cos(\alpha+\beta) = v(t-t_0)\frac{\cos(\alpha+\beta)}{\cos\alpha}. \tag{4}$$

Taking into account Eqs. (3) and (4) and the discussion associated with Fig. 1, Eqs. (1) and (2) can be rewritten as

$$c(t-t_0) = \lambda_0 + \frac{v(t-t_0)}{\cos\alpha}, \tag{5}$$

and

$$\lambda = v(t-t_0)\frac{\cos(\alpha+\beta)}{\cos\alpha} + c(t-t_0). \tag{6}$$

We express the term $(t - t_0)$ from Eq. (5) as

$$t - t_0 = \frac{\lambda_0}{c - \dfrac{v}{\cos\alpha}} \tag{7}$$

and make a substitution into Eq. (6) to obtain

$$\frac{\lambda}{\lambda_0} = \frac{c + v\dfrac{\cos(\alpha + \beta)}{\cos \alpha}}{c - \dfrac{v}{\cos \alpha}}. \tag{8}$$

The result can be further simplified if we eliminate the angle of reflection $\beta$ by using the trigonometric identities

$$\cos(\alpha + \beta) = \cos \alpha \cos \beta - \sin \alpha \sin \beta \tag{9}$$

and

$$\sin \beta = \sqrt{1 - \cos^2 \beta}, \tag{10}$$

and the formula

$$\cos \beta = \frac{-2\dfrac{v}{c} + \left(1 + \dfrac{v^2}{c^2}\right)\cos \alpha}{1 - 2\dfrac{v}{c}\cos \alpha + \dfrac{v^2}{c^2}} \tag{11}$$

that relates the angle of incidence $\alpha$ and the angle of reflection $\beta$ [3,6,7]. After some simple algebra, Eq. (8) can be transformed into

$$\frac{\lambda}{\lambda_0} = \frac{1 - \dfrac{v^2}{c^2}}{1 - 2\dfrac{v}{c}\cos \alpha + \dfrac{v^2}{c^2}}. \tag{12}$$

Hence, the formula that connects the frequencies $f_0$ and $f$ of the incident and the reflected light, respectively, states

$$f = f_0 \frac{1 - 2\frac{v}{c}\cos\alpha + \frac{v^2}{c^2}}{1 - \frac{v^2}{c^2}}, \qquad (13)$$

where we have taken into account that $\lambda_0 = c/f_0$ and $\lambda = c/f$. The result coincides with Einstein's formula obtained with Lorentz transformations [2].

The Doppler effect from a uniformly moving mirror is usually not considered in the standard textbooks on special relativity, although its origin belongs to the foundations of the theory. We hope that the method of derivation in this paper will bring this important issue into the undergraduate classroom and will enthuse the student due to its simplicity to arrive at the same Doppler formula.

The author is indebted to Professor Dan Censor (Ben-Gurion University of the Negev, Beer Sheva, Israel) for very helpful comments and suggestions that substantially improved the quality of the presentation.

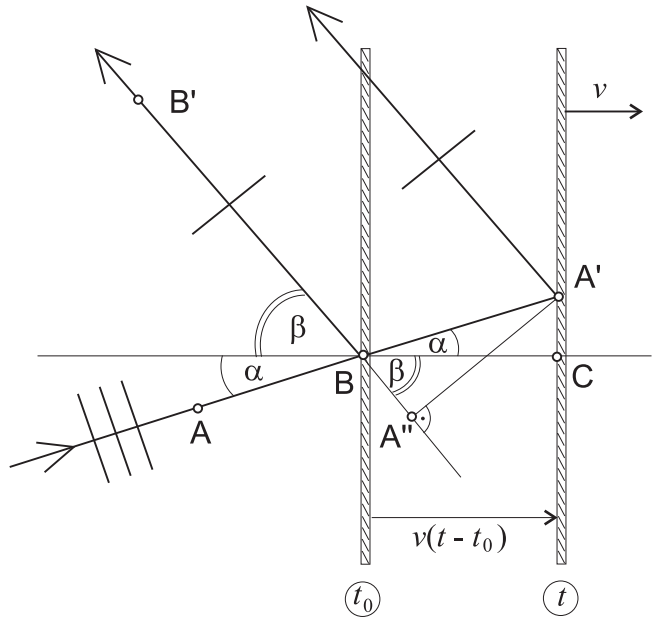

**Fig. 1.** The derivation of the formula for the Doppler shift from a uniformly moving vertical mirror. The mirror is moving at a constant velocity *v* to the right. The planes of the incident and the reflected wavefronts are perpendicular to their paths.